\documentclass{article}

\PassOptionsToPackage{numbers, compress}{natbib}
\usepackage[final]{neurips_2023}

\usepackage[utf8]{inputenc} 
\usepackage[T1]{fontenc}    

\usepackage{hyperref}       
\usepackage{url}            
\usepackage{booktabs}       
\usepackage{amsfonts}       
\usepackage{nicefrac}       
\usepackage{microtype}      
\usepackage{xcolor}         

\usepackage{enumitem}

\usepackage{graphicx}
\usepackage{float}

\setlist[itemize]{leftmargin=*}

\newcommand{\ie}{\textit{i.e.,}}
\newcommand{\eg}{\textit{e.g.,}}

\title{Assessing AI Impact Assessments: A Classroom Study}

\author{%
  Nari Johnson\\
  Carnegie Mellon University\\
  \texttt{narij@andrew.cmu.edu} \\
  \And
  Hoda Heidari \\
  Carnegie Mellon University \\
  \texttt{hheidari@andrew.cmu.edu} \\
}
\usepackage{enumitem}
\setlist[itemize]{leftmargin=*}

\begin{document}

\maketitle

\begin{abstract}
Artificial Intelligence Impact Assessments (``AIIAs''), a family of tools that provide structured processes to imagine the possible impacts of a proposed AI system, have become an increasingly popular proposal to govern AI systems.
Recent efforts from government or private-sector organizations have proposed many diverse instantiations of AIIAs, which take a variety of forms ranging from open-ended questionnaires to graded score-cards.
However, to date that has been limited evaluation of existing AIIA instruments.
We conduct a classroom study ($N = 38$) at a large research-intensive university (R1) in an elective course focused on the societal and ethical implications of AI.
We assign students to different organizational roles (\eg{} an ML scientist or product manager) and ask participant teams to complete one of three existing AI impact assessments for one of two imagined generative AI systems.
In our thematic analysis of participants' responses to pre- and post-activity questionnaires, we find preliminary evidence that impact assessments can influence participants' perceptions of the potential risks of generative AI systems, and the level of responsibility held by AI experts in addressing potential harm.
We also discover a consistent set of limitations shared by several existing AIIA instruments, which we group into concerns about their \emph{format} and \emph{content}, as well as the \emph{feasibility} and \emph{effectiveness} of the activity in foreseeing and mitigating potential harms.
Drawing on the findings of this study, we provide recommendations for future work on developing and validating AIIAs.
\end{abstract}

\section{Introduction}

One cannot understate the impact that Artificial Intelligence (AI) systems have had on society, from organizing our social media feeds \citep{eslami2014always} to influencing how we find work \citep{Raghavan_2020, akpinar2022longterm}, or informing critical decisions about our lives, such as whether we will receive a loan \citep{verma2022counterfactual} or be granted parole \cite{berk2017fairness}.
However, as an increasing number of incidents have demonstrated AI systems' potential to cause harm \citep{mcgregor2020preventing, shelby2023sociotechnical}, policymakers, regulators, and corporations face a new sense of urgency to govern AI.
One proposal which has gained traction among regulators \citep{calvi2023enhancing, oduro2022obligations, halim2023vectors} is AI Impact Assessements, or ``AIIA''s.
Inspired by the long history of impact assessments in other scientific or policy domains (\eg{} environmental impact assessments) \citep{ortolano1995environmental}, AIIAs provide structured processes for organizations to consider the implications to people and their environment of a proposed AI system.
Recent efforts have proposed many diverse instantiations of impact assessments for AI tools (surveyed by \citep{stahl2023systematic}) which range in form from open-ended questionnaires to graded score-cards, some of which are already in-use or mandatory under existing governance regimes \citep{canadianaia,microsoft2022rai, oduro2022obligations}.
However, to date there has been limited empirical evaluation of existing AIIA instruments \citep{moss2021assembling}.

\textbf{Contributions.} 
In this work, we conduct a \emph{preliminary classroom study} to explore the usability and effectiveness of existing AIIA templates.
We design a role-playing activity where we assigned $N = 38$ students enrolled in an elective course focused on the societal and ethical implications of AI, to different organizational roles, \eg{} an ML scientist or product manager.
We then asked participant teams to complete one of the three prominent AI impact assessment instruments for one of two hypothetical generative AI systems: a general-purpose video chatbot, and a video chatbot fine-tuned to conduct initial screening interviews with job candidates.
We surveyed participants immediately before and after the group activity to understand how completing an AIIA influenced their thinking about the potential impacts of generative AI, and collected their feedback on each template.

Our thematic analysis of students' responses provides preliminary evidence that existing AIIA instruments \emph{can} influence respondents' perceptions of the potential risks of generative AI systems.
After completing an AIIA, several students reported an \emph{increased level} of concern and perceived level of responsibility as machine learning experts, and collectively reported a more comprehensive and actionable set of potential issues with their product.
Thus, existing AIIAs have the potential to promote the imagination and mitigation of potential harms.
However, our analysis of students' critiques reveals a consistent set of limitations shared by prior AIIA instruments, which we group into concerns about their \emph{format} and \emph{content}, as well as the \emph{feasibility} and \emph{effectiveness} of the activity in foreseeing and mitigating potential harms.
Drawing on the findings of this study, we provide several recommendations for future work to co-design new and improved AIIA instruments.

\section{Background}
\paragraph{Related Work} Following \citep{stahl2023systematic}, we adopt a broad definition of an AI impact assessment as a structured process to help stakeholders understand the implications of a proposed AI system.\footnote{Several instruments that we consider use the term ``algorithmic impact assessment'' (AIA) and are developed to assess specific types of automated systems. We include these instruments because of the wide overlap in their scope and applicability with existing AI impact assessments and AI products.}
While many related impact assessments (such as privacy \citep{ftc2023privacy} or data protection \citep{demetzou2019dpia, ivanova2020data} IAs) may be relevant to AI systems, we focus our study on instruments designed specifically to consider the impact of introducing AI.
Existing AIIA instruments vary widely along several critical dimensions, including their articulated scope, purpose, form, and content \citep{stahl2023systematic}. 
Existing instruments propose a wide variety of different workflows, at varying levels of specificity.
While all AIIA instruments outline a process for organizations to follow \citep{reisman2018algorithmic}, only a subset have a clear set of expected deliverables from their processes \citep{canadianaia, lovelace2022aia, microsoft2022rai}.
Many such instruments are structured as questionnaires designed to facilitate reflection about societal impact, where the AIIA is ``completed'' when all questions have been answered.
Some AIIA frameworks also propose processes by which the AIIA is then \emph{reviewed}, \eg{} public agencies should solicit public comments on their AIIA before moving forward with the proposed system \citep{reisman2018algorithmic}.
Given the limitations of conducting a classroom study, we focus our research on characterizing challenges faced by assessors when they are asked to fill out an AIIA questionnaire, and how the process of completing an AIIA affects their thinking about potential impacts of AI.
We note that developing an effective questionnaire (the focus of this work) is just one important step of assembling a functional and robust AIIA regime \citep{moss2021assembling, selbst2021institutional, costanza2022who, mäntymäki2023putting}.

In response to the rising popularity of AIIAs as a potential accountability mechanism for AI, a growing body of scholarship has proposed evaluative criteria or studies to assess the efficacy of AIIA instruments \citep{moss2021assembling, nuno2021ai, watkins2021governing, selbst2021institutional, costanza2022who}.
An increasing number of studies have run empirical validations of AIIAs, such as OpenLoop's experiment where researchers partnered with 10 European AI companies to co-design a new AIIA instrument \citep{nuno2021ai}.
Our research contributes to this growing and timely body of work that aims to critically evaluate and characterize potential limitations of AIIAs.

\paragraph{AIIA Instruments}  
Our classroom study was inspired by our simple observation that many of the most prominent and well-known AIIA instruments at the time which we ran our study were released without any publicly available evaluation of their usability or effectiveness.\footnote{Of the three templates we selected, two (the Canadian ADM and Microsoft) state that the template was revised through consultation with relevant stakeholders.  However, to our knowledge the details of this process are not publicly available.}
We selected three different instruments pictured in Figure \ref{fig:templates} (released by the US CIO Council, Canadian Treasury Board, and Microsoft) that past work has referenced as mature and well-known instantiations of a general, domain-agnostic AIIA \citep{moss2021assembling}, and that also differed in interesting ways.
We provide a detailed summary of the form and content (\eg{} what types of impact users were prompted to consider) of each AIIA instrument in Appendix \ref{apdx:instrument-overview}. 
All three instruments were at some point actively in-use within an existing AI governance regime, \eg{} all Canadian executive agencies must complete the Canadian AIIA for any proposed AI system \citep{canadianaia} and all Microsoft AI products were required to be assessed by AIIA \citep{microsoft2022rai}.
We invite the reader to explore each instrument online.\footnote{Students accessed each instrument online on March 30, 2023.  They accessed version 0.10.0 of the Canadian AIA (released on March 26th, \href{https://github.com/canada-ca/aia-eia-js/tree/98d4499d6b86b33abbda36ecdec291e6e54e6766}{source code}, \href{https://www.canada.ca/en/government/system/digital-government/digital-government-innovations/responsible-use-ai/algorithmic-impact-assessment.html}{link}), the alpha version of the US CIO AIA (\href{https://www.cio.gov/aia-eia-js/}{link}), and the June 2022 release of Microsoft's Responsible AI Impact Assessment (\href{https://blogs.microsoft.com/wp-content/uploads/prod/sites/5/2022/06/Microsoft-RAI-Impact-Assessment-Template.pdf}{link}).}

\section{Study Design}
We conducted an in-class research activity where we assigned students enrolled in an elective course to different organizational roles, and asked them to work together in teams to complete an AIIA.  
The study was approved by an Institutional Review Board (IRB) process and conducted synchronously on Zoom in a single 80-minute class session in March 2023.
We provide further details on our study design and text of all study materials in Appendix \ref{apdx:materials}, and summarize key details in this section.

\textbf{Activity overview.}  Students were assigned at random to teams, where each student within the team role-played an assigned organizational role to complete an existing AIIA for an imagined product scenario.

\textbf{Organizational roles.} Each student was assigned at random to a team of $3$ or $4$ students.
    Within each team, each participant was assigned to one of three roles: (1) a \emph{machine learning scientist} representing the team in charge of design or development, (2) a \emph{product manager} in charge of making the project economically successful, and (3) a \emph{user representative} meant to represent the interests of potential users of the product.
    Each team had at least one participant assigned to all three roles, and teams with four participants had two user representatives.

\textbf{Scenarios.} We designed $2$ product scenarios, and assigned half of the teams to each one. 
Both scenarios told participants to role-play that they are employed by a private firm that has developed a \emph{generative AI model} that is capable of powering a hyper-realistic video-chat agent.
While the AI model in both scenarios was described to have the same capabilities, the two scenarios differed in the \emph{level of specificity} of the system's intended use.
In one scenario, individual \emph{users} of the AI product could ``specify the characteristics of the agent they would like to converse with'', like other general-purpose products, \eg{} ChatGPT.
In the other scenario, the company was specifically working on a product that would use the agent to conduct initial job screening interviews with candidates.
We hypothesize that existing AIIA tools may be less useful for general-purpose technologies, \ie{} that teams tasked to complete the AIIA for a product where the \emph{user specifies the end use} may encounter more difficulties than teams who are given a more well-specified end use of screening candidates.
We provide the complete scenario text shown to participants in Appendix \ref{apdx:scenario-text}, which we wrote to provide students with the level of detail that would typically be available during the product ideation stage, \ie{} before any particular model has been developed.

\textbf{Study population.}
Our study population consisted of $38$ students at a large research-intensive university (R1)  enrolled in an elective course on the societal and ethical considerations of AI.
All participants had self-reported intermediate knowledge of machine learning, as introductory machine learning was listed as a prerequisite for the course.
Students are future technologists and practitioners who may someday be asked to complete an AIIA, and as such we believe that their feedback is valuable in creating an AIIA template that is broadly accessible.
There are several important limitations when using a classroom role-playing exercise to evaluate a proposed AIIA workflow: for example, students do not have situated expertise that practitioners would have (\eg{} in a hiring scenario), and we did not provide them context on existing governance or accountability structures within their simulated organizations.
Thus, we believe our study, while informative, is no replacement to running grounded empirical evaluations of proposed AI systems within real organizational contexts.

\textbf{Study procedure \& data collection.} The course instructor began the study by introducing the study task, team assignments, and scenarios.
Before beginning the role-playing exercise, students individually completed a \emph{pre-questionnaire} form.
Once they completed the pre-questionnaire, students entered break-out rooms with their team and begin the role-playing activity (\ie{} fill out the assigned impact assessment template).
Students were specifically told to \emph{think-aloud} by verbalizing all of their thoughts as they worked together to complete the impact assessment.
Once they completed the impact assessment (or the course time ended) students individually completed a \emph{post-questionnaire} form.
We provide the complete text of the pre- and post-questionnaires in Appendices \ref{apdx:pre-q} and \ref{apdx:post-q}.

\subsection{Research Questions}

Our first set of research questions compare students' answers to the same set of questions (copied verbatim from the pre- to post-questionnaire) to examine if, and how, completing an AIIA changed their thinking about both the potential implications of the imagined AI product, and more broadly about generative AI.

\begin{itemize}
    \item \textbf{RQ1}: \emph{Imagining potential harms}. How does completing an AIIA affect the issues and values that students believe are the most urgent to address for their product? 
    \item \textbf{RQ2}: \emph{Sentiment towards generative AI}. How does completing an AIIA affect students' self-reported excitement and concern about potential uses of generative AI (beyond the presented scenario)? 
    \item \textbf{RQ3}: \emph{Responsibility of ML experts}. How does completing an AIIA affect students' perceived relative level of responsibility held by machine learning experts in addressing potential harms?
\end{itemize}

Our final research question analyzes students' reflections about their AIIA after completing the exercise:

\begin{itemize}
    \item \textbf{RQ4}: \emph{Characterizing limitations of \& opportunities for AIIA instruments}. What concerns or critiques do students have of existing AIIA instruments? 
    What considerations should those designing improved AIIA instruments keep in mind?
\end{itemize}

For all of our research questions, we also explore differences in participants' responses across treatments (\ie{} stratified by AIIA instrument or scenario condition). 
For our quantitative analyses (RQ2 and RQ3), we report differences in measures (such as the average self-reported excitement score across participants) across conditions. 
For our thematic analyses, we highlight what we believe are particularly informative differences across instruments or scenarios.

\paragraph{Data analysis}
In our quantitative data analysis, we did not run statistical significance tests. Instead, we opted to  report summary statistics only and compare them with students' responses to closed-form questions. Due to our small sample size and the nature of the study, our numerical reports should be understood as suggestive evidence (and not statistically significant claims) to contextualize the qualitative findings and inform hypothesis making for future larger-scale studies, conducted with a more representative sample of the target population. 
To analyze the free response questions, we adopted a thematic analysis approach where we qualitatively coded participants' responses in a shared coding session.
We conducted a bottom-up affinity diagramming process \citep{beyer1999contextual} to identify higher-level groups.
We then compared the frequency of each code across participants' experimental conditions, \ie{} across the template or scenario conditions, to understand if themes were shared across or unique to a particular template or scenario condition.

\section{Findings}
In this section, we summarize key findings from our analysis of participants' pre- and post-questionnaires, organized by research question.
We contextualize our thematic analyses using quotes from participants' responses, and refer to each participant using their four-digit ID number.

\subsection{Imagining Possible Impacts (RQ1)}
\begin{figure}[H]
\centering
\includegraphics[width=1\linewidth]{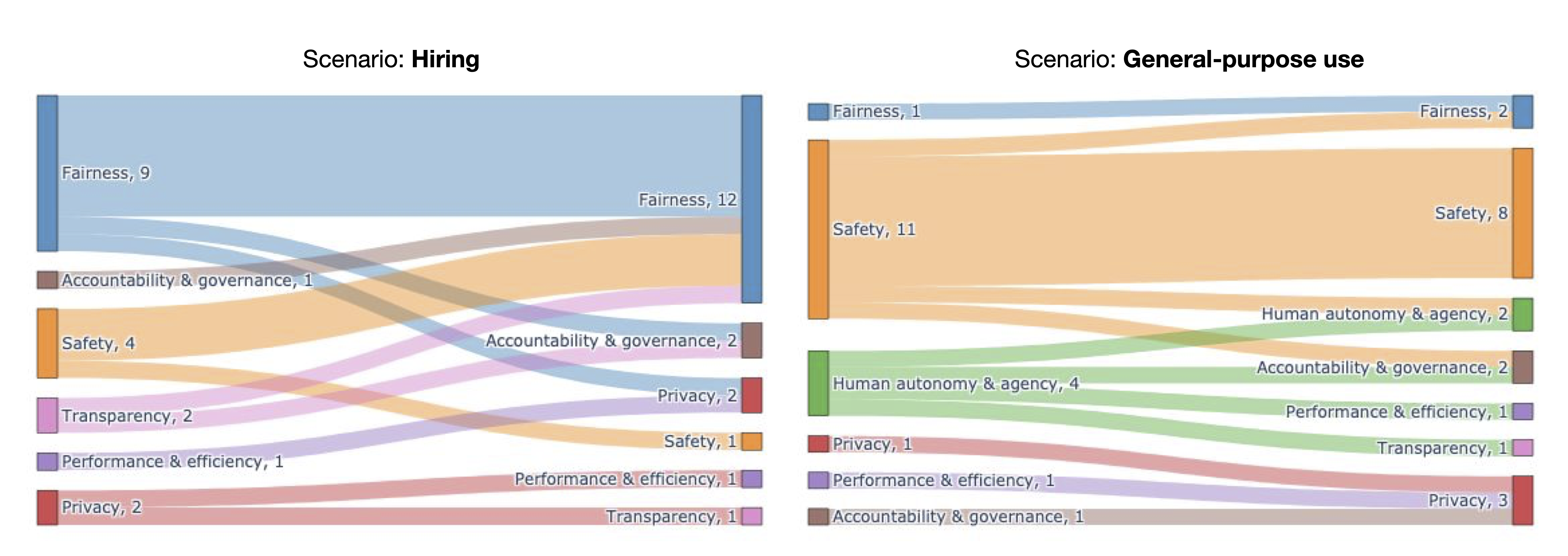}
\caption{Sankey diagram to visualize students' changes in beliefs (\ie{} their answers to the question, ``\emph{Which one of the following broad categories of values do you believe is most urgent to address for your product?}'') before (left) and after (right) completing the AIIA.  We group students' responses by the scenario (hiring vs. general-purpose use) that their team was assigned. 
We observed that students responses differed depending on their scenario: students assigned to complete the AIIA for the hiring product were significantly more likely to rate ``\emph{Fairness}'' as most urgent, while a much larger number of students assigned to the general-purpose use scenario chose ``\emph{Safety}'' or ``\emph{Human autonomy \& agency}''.
See Appendix \ref{apdx:value-categories} for complete definitions of each value.
}
\label{fig:seven-values}
\end{figure}

Before and after they completed the AIIA, we asked students to report which one out of seven possible categories of values they believed ``\emph{is most urgent to address for [their] product}'' (see Appendix \ref{apdx:value-categories} for the definitions of each value).
We visualize students' changes in beliefs before and after completing the AIIA in Figure \ref{fig:seven-values}.
We observed that students' responses varied significantly based on which scenario they were assigned: a larger number of students tasked to complete the AIIA for the hiring product rated ``\emph{Fairness}'' as most urgent, while a much larger number of students assigned to the general-purpose use scenario chose ``\emph{Safety}'' or ``\emph{Human autonomy \& agency}''.
Many students also changed their response after exposure to the AIIA: 50\% of all students chose a different value in their post-questionnaire than their original choice in the pre-questionnaire. 

In addition to selecting the single most urgent value, we also asked students to ``\emph{outline [in free text] the top three issues you believe the development team of the AI product assigned to your team should carefully address}''.
Before completing their AIIA, students reported a wide range of potential issues for both the hiring and general-use products in their pre-questionnaires.
Common responses shared by students across \emph{both} scenarios included concerns about the accuracy, effectiveness, and functionality of the system; data privacy, security, and collection without users' consent; data quality or dataset bias; potential to be ``biased'' or ``unfair'' against protected groups; and different dimensions of transparency \citep{vassei2023ai}, \eg{} whether the AI is \emph{explainable} and whether the user is notified that they are interacting with an AI.

Beyond these shared concerns, students also \textbf{identified concerns that were bespoke to their given scenario}.
Students tasked with the hiring scenario reported concerns about gaming, as the AI product may introduce ``\emph{more opportunities for the candidate to cheat} (P1462), or may ``\emph{be gamed if candidates state relevant but uninformative key words}'' (P1541).
Students tasked with the general-purpose use scenario listed concerns about its potential for malicious use, including creating deepfakes (\eg{} ``\emph{stealing someone's identity}'', P1153), or violent, explicit, or toxic content (P1384, P1063, P1816).
Several students also expressed concern about users' potential over-reliance or persuasion by faulty AI: ``\emph{the AI may play a role in giving users information, so it is important that the AI is as unbiased and factual as possible, to not mislead the user with false or radical ideas}'' (P1958).

When we compared individual students' responses to the same question \emph{after} they completed the AIIA, we observed two interesting behaviors.
First, \textbf{different students that were presented with the same instrument articulated a \emph{similar} set of \emph{new} concerns (that were not listed in their pre-questionnaires) after completing the AIIA}.
For example, $5$ students who completed the Microsoft template (and no other students) listed stereotyping as one of their top concerns after completing their AIIA, and several students that completed the US CIO AIIA  expressed concern about how a general-purpose conversation agent would affect children.
Notably, in both of these cases, the students' AIIA templates explicitly prompted them to consider these specific impacts (\eg{} the US CIO AIIA was the only template to include a question about the system's ``\emph{impact on children under the age of 18}''). 
This finding illustrates how AIIA instruments can inform students' awareness of possible harms that they may have previously not prioritized or considered.
Further, that students prioritized potential issues that were \emph{unique to their AIIA instrument} demonstrates how the \emph{content} of each AIIA (\ie{} which particular harms are emphasized or excluded) has the potential to influence users.

Second, we observed that \textbf{existing AIIAs facilitated reflection on new types of impact that \emph{no} student had listed as being important in their pre-questionnaires}.
After completing their AIIA, several students listed that they believed the system's \emph{broader impacts} to society and individuals whom are not direct users of the product, such as ``\emph{job displacement}'' (P1387, Microsoft) and ``\emph{economic effects of automation (who and what will be replaced)}'' (P1081, Canadian), were most important to address.
Students also emphasized the importance of developing \emph{processes} (such as audit trails or procedural fairness) to facilitate governance best practices, such as developing a ``\emph{rigorous logging and audit trail [...] to capture model failures and address potential user harm as quickly as possible}'' (P1359, Canadian) and ``\emph{ensuring that the model is auditable}'' (P1384, Canadian).
We also observed that more students listed preliminary ideas or explicit actions to \emph{mitigate} potential negative impacts of the AI system (rather than simply naming potential harms) after completing the AIIA: for example, one participant named ``\emph{[implementing] ways to check whether [generated] recordings are real or AI-generated}'' as an important issue for the general-purpose use product.

\subsection{Shifting Mindsets: Excitement, Concern, and Responsibility (RQs 2 \& 3)}

\paragraph{Excitement and concern for general-purpose technology}  
Overall, most students entered into the AIIA exercise with high levels of both excitement and concern about potential uses of foundational generative AI models.
While seemingly contradictory, being simultaneously excited and concerned is a common sentiment voiced by many politicians, experts, and regulators who share great concern about potential harms of AI systems, absent measures for accountability and governance \citep{ostp2022blueprint}.
After completing their AIIA, the class' average excited rating slightly \emph{decreased} (-$0.16$ points) from $4.11$ to $3.94$ (where a rating of $5$ indicates ``very excited''), and concerned rating \emph{increased} (+$0.32$ points) from $4.00$ to $4.32$ (where a rating of $5$ indicates ``very concerned'').
A greater proportion of students changed their concernedness rating after completing the AIIA: only 24\% of students changed their excitedness rating, while 50\% of students changed their concern rating.
Of the students that changed their rating, $7$ out of $9$ students became \emph{less} excited, and $15$ out of $19$ students became \emph{more} concerned.
We observed differences in how students' self-reported excitement and concern changed for different instruments, and different scenarios (complete results in Appendix \ref{apdx:results-extended}), where students who were given the general-purpose use scenario and US CIO or Microsoft AIIAs had relatively larger increases in their concern.
We hypothesize that there was a slightly greater effect on participants' concern because existing impact assessments are primarily focused on articulating potential \emph{negative} rather than positive impacts of proposed systems.

\paragraph{Responsibility of ML experts} To survey students' perceptions of \emph{who} should be held accountable to mitigate potential impacts, we asked them to rate ``\emph{the relative level of responsibility you imagine for ML experts (compared to other stakeholders) to address [the issues they identified]}''. 
The class' average rating increased ($+0.26$) from a score of $4.24$ to $4.50$ after completing the AIIA.
$20$ (52\% of all) students changed their response in their post-questionnaire, the majority ($15$) of whom reported an \emph{increased} relative level of responsibility for ML experts.
The extent to which students' ratings increased varied across AIIA instruments, where the Canadian AIIA resulted in the largest average increase (+$0.42$), which we speculate may be because the Canadian AIIA had the most detailed and thorough mitigation sections (summarized in Appendix \ref{apdx:instrument-overview}).
We present extended results of how ratings varied across conditions in Appendix \ref{apdx:resp-results}.

\subsection{Limitations of Existing Impact Assessment Instruments (RQ4)}

We highlight interesting themes that emerged in students' reflections of the limitations of each AIIA, \ie{} their responses to the question: ``\emph{What do you think are the major limitations of the AIA you completed for your team's product?}'' (RQ4).
We found that students both listed a consistent set of limitations that were shared by all three AIIA instruments, and also identified limitations specific to individual instruments.
We further group students' responses into concerns about instruments' \emph{format}, \emph{content}, \emph{feasibility}, and \emph{effectiveness}.

\begin{figure}[H]
\centering
\includegraphics[width=0.7\linewidth]{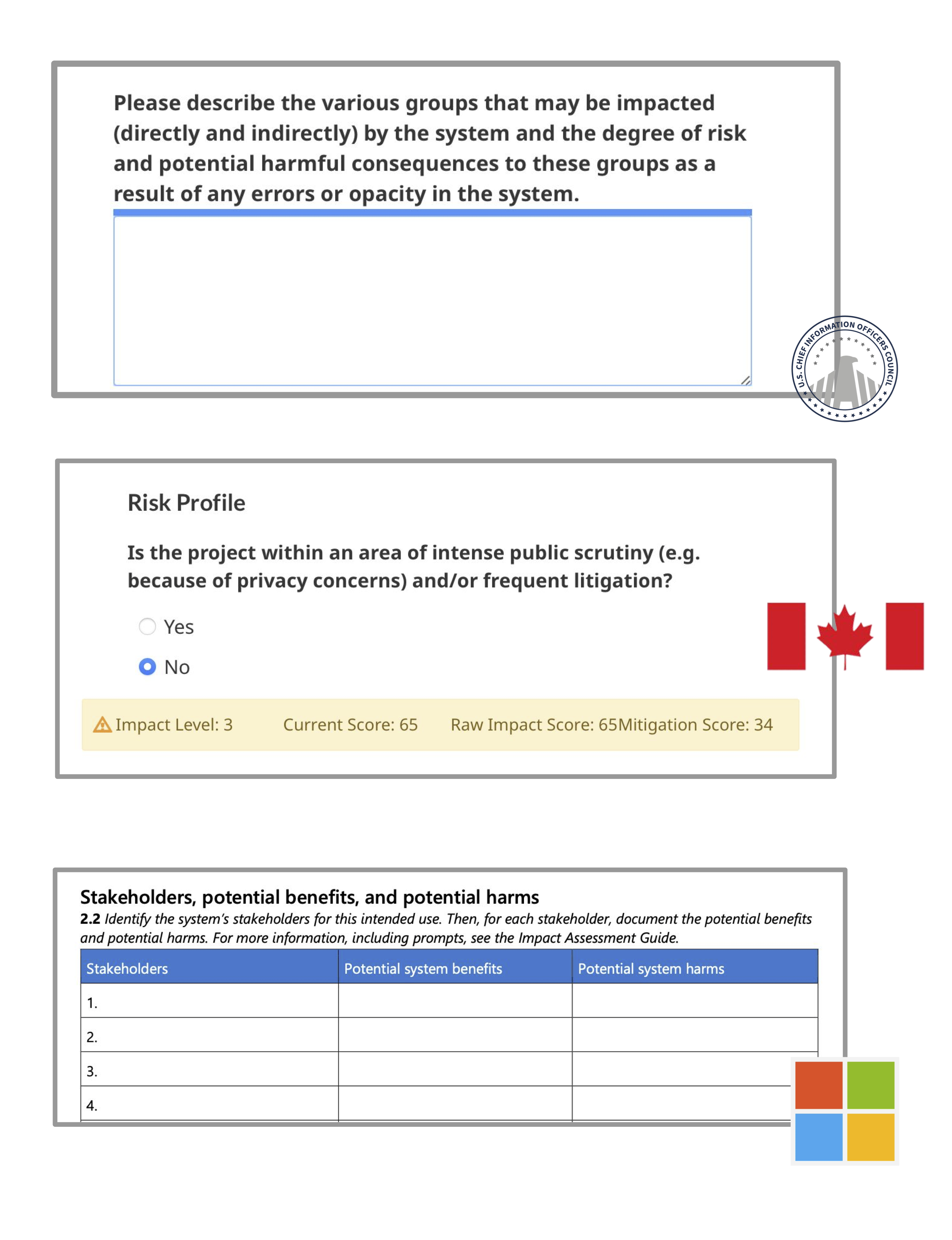}
\caption{Screenshots of a subset of questions that prompt users to imagine potential harms from the three AIIA instruments evaluated in our study. Both the US CIO (Top) and Canadian Treasury (Middle) templates share a similar responsive web format, containing mostly closed-form questions (such as the `Yes/No' questions pictured here) and the occasional free-response question.  In contrast, the Microsoft template (Bottom), which is a static PDF, consists mostly of free-response exercises (such as filling out the pictured table). 
The Canadian template (Middle) calculates a series of scores pictured in the yellow bar, which are used to determine mitigation steps required for the proposed AI system \citep{canadianaia}.
}
\label{fig:templates}
\end{figure}

\paragraph{Format.}  While each AIIA consisted of a series of questions that users must respond to, each AIIA instrument varied in its \emph{format} -- design decisions such as whether the AIIA is administered as a static form or dynamic website, or the expected response type of each question (\eg{} multiple choice or free response).
We summarize key findings that are either shared across instruments or bespoke to individual instruments:

\begin{itemize}
    \item \textbf{The AIIA has open-ended questions that grant freedom, yet require creativity}.  All three AIIA instruments included open-ended questions that elicited free text responses. 
    One student pointed out that due to this open-endedness, the quality of the final completed AIIA ``\emph{depends heavily on the reliability and creativity of the person/team filling it out}'' (P1709, Microsoft). Another student also noted that the instrument required them to make several subjective judgments because several terms were under-specified: ``\emph{it required a lot of qualitative assessments like those for safety and reliability and intelligibility that were not well-defined and could vary a lot depending on who fills out the AIIA and when}'' (P1750, Microsoft). 
    
    \item \textbf{The AIIA has close-ended questions that hide nuance}. Participants also critiqued questions that restricted responses to a specific form, such as answers to Yes/No questions about the AI system, or checklists and multiple choice questions that asked users to sort the AI into relevant categories. One participant explicitly called attention to a specific section of their instrument (Microsoft Section 5.2, ``Goal Applicability''): ``\emph{the [...] section classifies the systems into only two classes (i.e., yes or no). This may not measure the systems well.}'' (P1816).

    The Canadian AIIA differed from the others in that it used users' responses to closed-ended questions to calculate a numeric ``raw impact score'' and ``risk score'', which were displayed at the bottom of the AIIA webpage (shown in Figure \ref{fig:templates}). These scores were then used to sort the AI into one of four ``impact levels'', which determine the mitigation steps that the system owners are required to take.
    However, many participants stated that they did not understand how the numeric scores were calculated (P1081), or found them difficult to interpret (P1384, P1153).
    One participant expressed a desire to better understand the method used to calculate the risk score: ``\emph{we were confused about how points corresponded to our answers. It may be more helpful to be able to walk-through (after a first draft is completed) and see what alternative choices would result in a lower score}'' (P1081).

\end{itemize}

\paragraph{Content.}  The three AIIA instruments had overlap, yet varied considerably in their \emph{content}: the types of impacts and mitigation steps that users were prompted to consider.  We provide a detailed comparison across instruments in Appendix \ref{apdx:instrument-overview}, and summarize participants' critiques below:

\begin{itemize}
    \item \textbf{The AIIA is missing types of impact}. Several participants noted that they believed their AIIA was not comprehensive enough. Participants wrote that their instrument's ``\emph{categorized limitations and harms}'' are ``\emph{non-extensive}'' (P1063, Canadian), or that their AIIA ``\emph{doesn't address broader issues beyond privacy and non-discrimination}'' (P1281, Microsoft).  Another participant explicitly noted types of impact that they believed should have been included in their AIIA: ``\emph{no question had us think critically about the accessibility of the system, or which subgroups/subpopulations could be disproportionately harmed if the model was to be deployed}'' (P1081, Canadian).\footnote{The Canadian AIIA has been updated to include additional questions regarding accessibility since the students participated in the study.}
    
    \item \textbf{The AIIA includes questions that may not be applicable to all AI systems}.  
    Some participants commented that not all questions may be relevant to all AI systems. One participant that completed the Microsoft AIIA found that ``\emph{some questions did not strictly apply to our product}'' (P1819).
    This calls into question whether it makes sense to assess all proposed AI systems with a common (versus more specialized) AIIA instrument, a tension that has been raised by past work \citep{moss2021assembling, nuno2021ai, stahl2023systematic}.
\end{itemize}

\paragraph{Feasibility.} While participants were instructed to complete the AIIA in a facilitated classroom exercise, several participants pointed out perceived or imagined obstacles that may prevent organizations from completing an AIIA in practice. We group these concerns as those related to the \emph{feasibility} of the AIIA, \ie{} what realistically must happen to effectively complete each questionnaire.

\begin{itemize}
    \item \textbf{Users completing an AIIA desire supervision}.  Several students listed a lack of supervision (\ie{} ``\emph{no oversight}'') as a limitation of the AIIA process (P1505, P1709).
    While all three instruments provided an email address that users could contact with questions or feedback on each instrument, only the Canadian AIIA listed named persons who could provide synchronous help and consultation.
    
    \item \textbf{Users completing an AIIA desire clear expectations and transparency about how their responses will be reviewed}.
    Some students expressed uncertainty or confusion about how their completed AIIA would be evaluated or reviewed (P1613, P1709), or were ``\emph{unsure if we answered the questions properly}'' (P1456).
    One student expressed concern specifically about how their open-ended responses would be evaluated: ``\emph{a lot of the questions were short answer, which makes sense since there's many different types of responses, but this sometimes made it unclear what should be answered}'' (P1613).

    \item \textbf{Users may lack information about the AI that is necessary for the AIIA at the time of its completion}.
    Study participants were given a short one-paragraph description of a potential AI system.
    However, many students reported that this  description did not provide enough information about the product to complete the AIIA (P1620, P1613, P1866, P1286, P1528).
    One participant commented that their AIIA ``\emph{seems to be designed for more fleshed out products rather than just ideas}'' (P1286, Microsoft). 
    Another participant's team ``\emph{didn't have a lot of context about the development of the AI system itself, so when answering these questions, we made a lot of assumptions about use cases and the development process}'' (P1613, US CIO).
    When participants opted against making restrictive assumptions, they noticed that existing instruments did not allow them to consider multiple downstream possibilities: 
    ``\emph{some of the questions where we need to select only a single option to be hard to choose from since the product might apply to multiple options provided (for example, the role of humans in the product)}'' (P1819, Microsoft).

    We note that participants' critiques may not only reflect potential limitations of the AIIA templates, but also of our study design (\eg{} that students lacked knowledge or context about the imagined scenario that a practitioner would have).
    The scenario text (pasted in Appendix \ref{apdx:scenario-text}) includes details of the proposed user interface but excludes implementation details such as the training dataset or model architecture, and was designed to simulate the information available in an AI services' early ideation stages.
    Thus, we hypothesize that many of the above critiques may still be relevant for real AI systems that are at a similarly early phase of their design.
    
    \item \textbf{The AIIA requires coordinating a team of stakeholders with interdisciplinary expertise}.  
    Several participants reported that they lacked relevant knowledge or the appropriate domain expertise necessary to complete the AIIA.
    One participant that completed the US CIO AIIA was confused by ``\emph{legality questions}'', writing, ``\emph{as technical members of the team, we were unfamiliar with the terminology used}'' (P1456).
    Another participant noted a ``\emph{lack of explanation on some of the terms [and] regulations}'' (P1696, Canadian).
    This finding points to the practical need to assemble a \emph{team} of stakeholders with appropriate interdisciplinary expertise, and identifying who is responsible for each piece.
    Existing AIIA interfaces presently do not include support for collaborating asynchronously across multiple devices.

\end{itemize}

\paragraph{Effectiveness.}  Participants reflected on whether the AIIA was effective at helping them imagine, and develop a plan of action to mitigate potential harmful impacts of the proposed AI.

\begin{itemize}
    \item \textbf{Potential harms \& mitigation steps are vague or under-specified}.
    Several participants reported that their AIIA used language that was ``\emph{not clear}'' (P1145, P1505), ``\emph{vague}'' (P1005, P1886), or ``\emph{difficult to understand}'' (P1620). 
    Unclear language throughout the AIIA posed challenges to helping users imagine and enumerate specific harms: ``\emph{many of the questions were vague and did not go over the harms that could be caused}'' (P1886, Canadian). 
    Some participants specifically called attention to their AIIA's suggested mitigation steps, which they felt were under-specified: participants wrote that Microsoft's AIIA ``\emph{doesn't provide specific guidelines on how to address risks}'' (P1281), ``\emph{didn't give us concrete directions of actions we need to take}'' (P1121), and ``\emph{while good at identifying harms, [is] not very helpful in suggestions for mitigation}'' (P1286).
    One participant noted how this under-specification could potentially be exploited in favor of the organization completing the AIIA, thus making it less effective as an accountability mechanism: ``\emph{many of the evaluations for mitigations were concerned with the existence of a procedure to handle something, but allowed this to be an in-house procedure and did not lay down any specifications as to what the procedure should consist of}'' (P1359, Canadian).

    \item \textbf{Existing AIIA instruments are less effective for general-purpose technologies}.  Across all three instruments, participants who were assigned to complete the AIIA for the general-purpose technology scenario reported a disconnect where parts of their AIIA did not seem applicable to general-purpose technologies.
    Participants that were shown the Canadian AIIA were especially more likely to note this disconnect (P1359, P1384, P1153, P1866, P1273), likely because the Canadian AIIA includes specific sections where the user must report the desired end use of the system (Section 7).
    Participants wrote that ``\emph{the questionnaire seemed to be designed for someone both creating a model and applying it to a given application domain. It didn't work as well for our case of creating a model that could be used in many domains}'' (P1359) and that ``\emph{the AIIA does not account for the difference in scope of the project vs the potential applications}'' (P1153).  Beyond critiques of existing instruments, one participant found imagining the potential impacts of a general-purpose system to be more challenging: ``\emph{I think the major limitation is that the description of the product was vague so depending on which industry the product was being applied to there could be all sorts of different issues that AIIA could not really pick up}'' (P1866).
\end{itemize}

\section{Discussion}
\vspace{-8pt}
Overall, our study provides preliminary evidence that exposure to existing AIIA templates \emph{can} influence respondents' perceptions about generative AI systems (RQ2, RQ3), and also affect the issues and values that they believe are most urgent to address (RQ1).
We analyzed students' critiques to identify four broad categories of concerns that limit the effectiveness of existing AIIA instruments (RQ4).
These gaps point to opportunities for future work on \textbf{co-designing and validating (1) new and improved AIIA instruments, and (2) processes and workflows around them} to ensure they are completed and acted on appropriately.

Our study sheds light on the several issues in existing \emph{instruments}. First, \emph{existing AIIA templates are incomplete}. 
They do not include a comprehensive set of likely harms, and the harms they identify are not adequately defined for the user. 
This finding points to several important directions for future work.
One direction is to conduct think-alound validation studies with a representative sample of stakeholders to ensure that key terms are interpreted correctly and uniformly.
Another is to devise a mechanism to encourage users to think about context-specific harms \citep{martelaro2020what}. 
Such approaches may explore how to meaningfully consult and involve impacted communities at the time of impact assessment \citep{sloane2020participation, moss2021assembling}.
Additionally, participants found that \emph{the user interface of existing AIIA templates are at times structured and restrictive, or too unstructured and open-ended}, hindering effective deliberation and imagination in both cases.
These challenges are not bespoke to AIIAs, but rather are common challenges that have been well-documented in related work on survey design \citep{converse1986survey}.
Future work can experiment with alternative interfaces and processes to strike a better balance between structured thinking and expressiveness.
We also encourage future work to engage with existing best practices from research in survey design and impact assessments across domains \citep{cameron2011facilitating, usaid2006ia}.


Perhaps more importantly, our work firmly establishes the need to embed AIIAs in the appropriate \emph{workflows and processes}. As pointed out by prior work \citep{moss2021assembling, selbst2021institutional}, many existing AIIAs don't clearly specify \emph{who} should participate in the impact assessment activity, and how different stakeholder groups should fill out the form individually or as a group. 
Further, existing AIIAs don't clearly specify \emph{when} the AIIA should be completed within the AI development lifecycle. 
For example, some existing AIIA templates ask the system owners to answer questions about the dataset or model architecture that will be deployed.
However, in practice these details are often unknown or subject to being changed at early phases of ideation, design, or development \citep{shankar2022operationalizing}.
Yet, past work has underscored the need for impact assessment and participation at the earliest possible phases of ideation, before too much investment into the system has been made \citep{reisman2018algorithmic}. 
Developing AIIA processes that are designed to be \emph{iterative and continued} across the stages of AI development \citep{raji2020closing}, and clarifying at what stage of development certain parts of an AIIA can be completed, are important directions for future work.

Finally, foreseeing the impacts of general-purpose AI technologies is particularly challenging due to the vast space of possible use cases and application domains \citep{rastogi2023supporting, shevlane2023model}. 
To ensure that risks are properly anticipated and mapped using existing templates, the intended use of the AI under assessment must be well-scoped and contextualized. 
Effective impact assessment for general-purpose AI may require designing new impact assessment specifically designed for general-purpose use.

While artificial intelligence impact assessments have been referenced in more and more proposals, few such proposals have been made concrete.
What exactly should an AIIA entail?
By eliciting students' feedback on existing AIIA templates, we outline preliminary criticisms and potential paths forward.
While existing AIIA templates are imperfect, we observed that they \emph{did} shape students' thinking on potential harms, and we believe are promising tools for imagining potential risks of AI systems.
Importantly, assessors' imagination of possible harms is just one of many articulated goals of AIIAs \citep{stahl2023systematic}, and is just one step within the larger process of creating an effective accountability regime \cite{selbst2021institutional, moss2021assembling, ainow2023algorithmic}, which involves stakeholders beyond the assessors (such as impacted communities and the general public).
More broadly, we hope that our work can inspire researchers and policymakers alike to conduct and share further empirical evaluations of AIIAs that assess their effectiveness for a wider set of stakeholders and goals.

\newpage

\begin{ack}
We thank our study participants who made this research possible.
We also thank the reviewers of the NeurIPS 2023 Regulatable ML workshop for their feedback.
H. Heidari and N. Johnson acknowledge support from NSF (IIS2040929 and IIS2229881) and PwC (through the Digital Transformation and Innovation Center at CMU).
Any opinions, findings, conclusions, or recommendations expressed in this material are those of the authors and do not reflect the views of the National Science Foundation and other funding agencies.
\end{ack}

\bibliographystyle{plain}
\bibliography{ref}

\begin{thebibliography}{10}

\bibitem{akpinar2022longterm}
Nil-Jana Akpinar, Cyrus DiCiccio, Preetam Nandy, and Kinjal Basu.
\newblock Long-term dynamics of fairness intervention in connection recommender systems, 2022.

\bibitem{berk2017fairness}
Richard Berk, Hoda Heidari, Shahin Jabbari, Michael Kearns, and Aaron Roth.
\newblock Fairness in criminal justice risk assessments: The state of the art, 2017.

\bibitem{beyer1999contextual}
Hugh Beyer and Karen Holtzblatt.
\newblock Contextual design.
\newblock {\em Interactions}, 6(1):32–42, jan 1999.

\bibitem{calvi2023enhancing}
Alessandra Calvi and Dimitris Kotzinos.
\newblock Enhancing ai fairness through impact assessment in the european union: A legal and computer science perspective.
\newblock In {\em Proceedings of the 2023 ACM Conference on Fairness, Accountability, and Transparency}, FAccT '23, page 1229–1245, New York, NY, USA, 2023. Association for Computing Machinery.

\bibitem{cameron2011facilitating}
Colleen Cameron, Sebanti Ghosh, and Susan~L. Eaton.
\newblock Facilitating communities in designing and using their own community health impact assessment tool.
\newblock {\em Environmental Impact Assessment Review}, 31(4):433--437, 2011.
\newblock Health Impact Assessment in the Asia Pacific.

\bibitem{ftc2023privacy}
The US Federal~Trade Comission.
\newblock Federal trade commission privacy impact assessments, 2023.

\bibitem{converse1986survey}
Jean~M. Converse and Stanley Presser.
\newblock {\em Survey questions : handcrafting the standardized questionnaire}.
\newblock Sage Publications, 1986.

\bibitem{costanza2022who}
Sasha Costanza-Chock, Inioluwa~Deborah Raji, and Joy Buolamwini.
\newblock Who audits the auditors? recommendations from a field scan of the algorithmic auditing ecosystem.
\newblock In {\em Proceedings of the 2022 ACM Conference on Fairness, Accountability, and Transparency}, FAccT '22, page 1571–1583, New York, NY, USA, 2022. Association for Computing Machinery.

\bibitem{demetzou2019dpia}
Katerina Demetzou.
\newblock Data protection impact assessment: A tool for accountability and the unclarified concept of ‘high risk’ in the general data protection regulation.
\newblock {\em Computer Law \& Security Review}, 35(6):105342, 2019.

\bibitem{eslami2014always}
Motahhare Eslami, Aimee Rickman, Kristen Vaccaro, Amirhossein Aleyasen, Andy Vuong, Karrie Karahalios, Kevin Hamilton, and Christian Sandvig.
\newblock "i always assumed that i wasn't really that close to [her]": Reasoning about invisible algorithms in news feeds.
\newblock In {\em Proceedings of the 33rd Annual ACM Conference on Human Factors in Computing Systems}, CHI '15, page 153–162, New York, NY, USA, 2015. Association for Computing Machinery.

\bibitem{lovelace2022aia}
Lara Groves.
\newblock Algorithmic impact assessment: a case study in healthcare, 2022.

\bibitem{ainow2023algorithmic}
AI~Now Institute.
\newblock Algorithmic accountability: Moving beyond audits, 2023.

\bibitem{ivanova2020data}
Yordanka Ivanova.
\newblock The data protection impact assessment as a tool to enforce non-discriminatory ai.
\newblock {\em Springer Proceedings of the Annual Privacy Forum}, 2020.

\bibitem{jakesch2022how}
Maurice Jakesch, Zana Bu{\c{c}}inca, Saleema Amershi, and Alexandra Olteanu.
\newblock How different groups prioritize ethical values for responsible {AI}.
\newblock In {\em 2022 {ACM} Conference on Fairness, Accountability, and Transparency}. {ACM}, jun 2022.

\bibitem{martelaro2020what}
Nikolas Martelaro and Wendy Ju.
\newblock What could go wrong? exploring the downsides of autonomous vehicles.
\newblock In {\em 12th International Conference on Automotive User Interfaces and Interactive Vehicular Applications}, AutomotiveUI '20, page 99–101, New York, NY, USA, 2020. Association for Computing Machinery.

\bibitem{mcgregor2020preventing}
Sean McGregor.
\newblock Preventing repeated real world ai failures by cataloging incidents: The ai incident database, 2020.

\bibitem{microsoft2022rai}
Microsoft.
\newblock Microsoft responsible ai standard, v2 (general requirements), 2022.

\bibitem{moss2021assembling}
Emanuel Moss, Elizabeth~Anne Watkins, Ranjit Singh, Madeleine~Clare Elish, and Jacob Metcalf.
\newblock Assembling accountability: algorithmic impact assessment for the public interest, 2021.

\bibitem{mäntymäki2023putting}
Matti Mäntymäki, Matti Minkkinen, Teemu Birkstedt, and Mika Viljanen.
\newblock Putting ai ethics into practice: The hourglass model of organizational ai governance, 2023.

\bibitem{nuno2021ai}
Norberto Nuno, Gomes de~Andrade, and Verena Kontschieder.
\newblock Ai impact assessment: A policy prototyping experiment, 2021.

\bibitem{oduro2022obligations}
Serena Oduro, Emanuel Moss, and Jacob Metcalf.
\newblock Obligations to assess: Recent trends in ai accountability regulations.
\newblock {\em Patterns}, 3(11):100608, 2022.

\bibitem{canadianaia}
The~Government of~Canada.
\newblock Algorithmic impact assessment tool, 2023.

\bibitem{ostp2022blueprint}
The White House~Office of~Science and Technology Policy.
\newblock Blueprint for an ai bill of rights: A vision for protecting our civil rights in the algorithmic age, 2022.

\bibitem{ortolano1995environmental}
Leonard Ortolano and Anne Shepherd.
\newblock Environmental impact assessment: Challenges and opportunities.
\newblock {\em Impact Assessment}, 1995.

\bibitem{halim2023vectors}
Leonard Ortolano and Anne Shepherd.
\newblock Vectors of ai governance - juxtaposing the u.s. algorithmic accountability act of 2022 with the eu artificial intelligence act.
\newblock {\em Berkman Klein Center for Internet \& Society}, 2023.

\bibitem{Raghavan_2020}
Manish Raghavan, Solon Barocas, Jon Kleinberg, and Karen Levy.
\newblock Mitigating bias in algorithmic hiring.
\newblock In {\em Proceedings of the 2020 Conference on Fairness, Accountability, and Transparency}. {ACM}, jan 2020.

\bibitem{raji2020closing}
Inioluwa~Deborah Raji, Andrew Smart, Rebecca~N. White, Margaret Mitchell, Timnit Gebru, Ben Hutchinson, Jamila Smith-Loud, Daniel Theron, and Parker Barnes.
\newblock Closing the ai accountability gap: Defining an end-to-end framework for internal algorithmic auditing, 2020.

\bibitem{rastogi2023supporting}
Charvi Rastogi, Marco~Tulio Ribeiro, Nicholas King, and Saleema Amershi.
\newblock Supporting human-ai collaboration in auditing llms with llms, 2023.

\bibitem{reisman2018algorithmic}
Dillon Reisman, Jason Schultz, Kate Crawford, and Meredith Whittaker.
\newblock Algorithmic impact assessments: A practical framework for public agency, 2018.

\bibitem{scassa2020administrative}
Teresa Scassa.
\newblock Administrative law and the governance of automated decision-making: A critical look at canada’s directive on automated decision-making.
\newblock {\em SSRN Electronic Journal}, 01 2020.

\bibitem{selbst2021institutional}
Andrew~D. Selbst.
\newblock An institutional view of algorithmic impact assessments.
\newblock {\em 35 Harvard Journal of Law \& Technology 117}, 2021.

\bibitem{shankar2022operationalizing}
Shreya Shankar, Rolando Garcia, Joseph~M. Hellerstein, and Aditya~G. Parameswaran.
\newblock Operationalizing machine learning: An interview study, 2022.

\bibitem{shelby2023sociotechnical}
Renee Shelby, Shalaleh Rismani, Kathryn Henne, AJung Moon, Negar Rostamzadeh, Paul Nicholas, N'Mah Yilla, Jess Gallegos, Andrew Smart, Emilio Garcia, and Gurleen Virk.
\newblock Sociotechnical harms of algorithmic systems: Scoping a taxonomy for harm reduction, 2023.

\bibitem{shevlane2023model}
Toby Shevlane, Sebastian Farquhar, Ben Garfinkel, Mary Phuong, Jess Whittlestone, Jade Leung, Daniel Kokotajlo, Nahema Marchal, Markus Anderljung, Noam Kolt, Lewis Ho, Divya Siddarth, Shahar Avin, Will Hawkins, Been Kim, Iason Gabriel, Vijay Bolina, Jack Clark, Yoshua Bengio, Paul Christiano, and Allan Dafoe.
\newblock Model evaluation for extreme risks, 2023.

\bibitem{sloane2020participation}
Mona Sloane, Emanuel Moss, Olaitan Awomolo, and Laura Forlano.
\newblock Participation is not a design fix for machine learning, 2020.

\bibitem{stahl2023systematic}
Bernd~Carsten Stahl, Josephina Antoniou, Nitika Bhalla, Laurence Brooks, Philip Jansen, Blerta Lindqvist, Alexey Kirichenko, Samuel Marchal, Rowena Rodrigues, Nicole Santiago, Zuzanna Warso, and David Wright.
\newblock A systematic review of artificial intelligence impact assessments.
\newblock {\em Artificial Intelligence Review}, 2023.

\bibitem{usaid2006ia}
USAID.
\newblock Collecting and using data for impact assessment, 2006.

\bibitem{vassei2023ai}
Ramak~Molavi Vasse'i, Jesse McCrosky, and Mozilla Insights.
\newblock Ai transparency in practice, 2023.

\bibitem{verma2022counterfactual}
Sahil Verma, Varich Boonsanong, Minh Hoang, Keegan~E. Hines, John~P. Dickerson, and Chirag Shah.
\newblock Counterfactual explanations and algorithmic recourses for machine learning: A review, 2022.

\bibitem{watkins2021governing}
Elizabeth~Anne Watkins, Emanuel Moss, Jacob Metcalf, Ranjit Singh, and Madeleine~Clare Elish.
\newblock Governing algorithmic systems with impact assessments: Six observations.
\newblock AIES '21, page 1010–1022, New York, NY, USA, 2021. Association for Computing Machinery.

\end{thebibliography}

\newpage
\appendix

\section{Overview of AIIA Instruments}\label{apdx:instrument-overview}

\textbf{US CIO AIA}.  The US CIO AIA (Figure \ref{fig:templates}, Top) was developed to ``\emph{help [US] federal government agencies begin to assess risk associated with using automated decision systems}''. 
At the time the study was conducted, the AIA's webpage states that it is in ``alpha mode'', and is presently not mandatory. 
The AIA is administered dynamically online, and contains a mix of free-response and multiple choice questions organized into $12$ total sections. The AIA presently has no review process or named persons available to contact for assistance.  Information in the AIA is only stored locally on one's own computer, and upon completing the assessment the user can export their responses to PDF. 

The first eight sections of the AIA prompt users to report various design decisions that they have made in the development of the system and its intended use, such as the agency's motives for using automation (Section 1), the named people responsible for deploying the system (Section 2), details about the data source, data provenance, and data privacy (Sections 3, 4, 5) and model (Section 7).
Section 8 of the AIA, titled ``System Impact Assessment and Risk Profile'', asks the user to reflect on different types of possible impacts their system may have to individual rights or freedoms, individuals' health or well-being, the environment, and the economy.
The final four sections ask the user to self-report any plans that they have made to mitigate potential harms, including consulting with external stakeholders (Section 9), following data quality and provenance best-practices (Section 10), and maintaining an ``audit trail'' \citep{raji2020closing} (Section 12).

\textbf{Canadian Treasury Board AIA}. The Canadian Treasury Board AIA (Figure \ref{fig:templates}, Middle) was developed to support the Canadian Government's Directive on Automated Decision Making, which mandates that all government institutions subject to the directive are required to complete and publicly release results of the AIA (by uploading them to an Open Government portal) for any automated decision system \citep{scassa2020administrative}.
Like the US CIO AIA, the AIA is administered dynamically online, and contains a mix of free-response and multiple choice questions organized into $13$ sections.

The Canadian AIA notably differs from the others studied in that it scores participants responses to closed-form questions in real time.
Specifically, the website calculates a numeric ``raw impact score'' and ``mitigation score'' (out of $51$ and $34$ total points, respectively), which are combined to sort the system into one of four ``impact levels''.
The final impact level of the system determines the mitigation steps that the organization is required to take under the Directive on Automated Decision-Making \citep{canadianaia}.

While the content of the Canadian AIA is similar to that of the US CIO, the two templates only share a small handful of questions.
Like the US CIO AIA, the Canadian AIA begins by asking a series of descriptive questions about the AI system, such as the agency's motives for introducing automation (Section 2), a ``risk profile'' that assesses the ``stakes'' of the system, the named persons responsible for completing the AIA and building the system (Sections 1 and 4), and basic information about the system, such as its capabilities (Sections 5 and 6).

The Canadian AIA's single ``Impact Assessment'' section is more thorough than that of the US CIO, which in addition to asking users to reflect on the same set of specific impacts from that template (\eg{} rights and freedoms, health and well-being, etc.), also includes many other questions about the larger social context in which the algorithm is embedded, \eg{} ``\emph{Please describe the output produced by the system and any relevant information needed to interpret it in the context of the administrative decision}'' and, ``\emph{Will the system perform an assessment or other operation that would not otherwise be completed by a human?}''.

Similarly, the Canadian AIA's three mitigation sections contain a larger and more comprehensive set of questions than the US CIO AIA.  
The ``Data Quality'' section asks if users plan to undertake processes to assess dataset bias or follow other best-practices (\eg{} releasing training data), and also asks if the user plans to ``\emph{make information [associated with these processes] publicly available}''. 
The ``Procedural Fairness'' section asks about users' plans to implement best practices such as maintaining an audit trail of documentation \citep{raji2020closing}, eliciting user feedback, or enabling human override of decisions. 
The final ``Privacy'' section asks if the users intend to complete a separate privacy impact assessment and lists potential mitigations to reduce risk (\eg{} de-identifying data before it is stored).

\textbf{Microsoft Responsible AI Impact Assessment}. In June 2022, Microsoft's Responsible AI team publicly released a PDF ``Responsible AI Impact Assessment Template'' (\href{https://blogs.microsoft.com/wp-content/uploads/prod/sites/5/2022/06/Microsoft-RAI-Impact-Assessment-Template.pdf}{link}), and accompanying guide (\href{https://blogs.microsoft.com/wp-content/uploads/prod/sites/5/2022/06/Microsoft-RAI-Impact-Assessment-Guide.pdf}{link}) in a public blog post (\href{https://blogs.microsoft.com/on-the-issues/2022/06/21/microsofts-framework-for-building-ai-systems-responsibly/}{link}) that publicly announced Microsoft's Responsible AI Standard.
The Responsible AI Standard attempts to operationalize Microsoft's six AI principles to define \emph{product development requirements} for all AI products at Microsoft.
Goal A1 of v2 of the standard states that under the standard, all Microsoft AI systems are assessed by impact assessments, which are then reviewed ``\emph{according to your organization's compliance before development starts}''.

One important difference that distinguishes the Microsoft template is that it targets the process of AI \emph{development} (\eg{} with product teams and engineers), rather than grounded real-world use of AI (as was the case for the US CIO and Canadian templates, which are intended to be completed by public agencies who are planning to use AI). 
Another important difference is the form of the template: the majority of questions are free-response, and ask users to fill out tables (\eg{} a table where each row corresponds to a different stakeholder, and each column facilitates reflection on potential benefits and harms they may experience).
In our classroom study, we presented students with the template PDF only (which does not contain links or references to the accompanying guide).

The content of the template is organized into five sections:
\begin{enumerate}
    \item The ``System Information'' section asks descriptive questions about the system, such as its present stage in the ``AI lifecycle'', intended purpose and uses, features, and areas where the system may be deployed.

    \item The ``Intended uses'' section asks the user to assess the system's ``fitness for purpose'', name potential benefits and harms to named stakeholders, and identify specific types of stakeholders (\eg{} name the person who will be responsible for overseeing the system post-deplyoment). 
    The section also has a section on ``fairness considerations'', that prompts the user to identify any ``demographic groups'' that may require fairness considerations. 
    The section also contains questions on technology readiness, task complexity, the role of humans in the system, and deployment environment.
    
    \item The ``Adverse Impact'' section prompts the user to reflect on potential uses beyond intended use, such as restricted use, unsupported use, sensitive use, or intentional/un-intentional mis-use.  The section also asks the user to report known limitations of the system and to imagine the impact of failure on stakeholders.

    \item The ``Data Requirements'' section asks the user to assess existing data requirements that may apply to the system and the suitability of available training datasets.

    \item The ``Summary of Impact'' section asks the user to ``\emph{describe initial ideas for mitigations}'' for the potential harms that they identified earlier in the IA. It also asks users to check a box for whether each of Microsoft's ``Responsible AI Goals'' (\ie{} product requirements grouped into accountability, transparency, fairness, reliability \& safety, privacy \& security, and inclusiveness) applies to their system. 
\end{enumerate}

\newpage
\section{Study Materials}\label{apdx:materials}

We include the complete text (shown to study participants) of our study materials below.

\subsection{Scenarios}\label{apdx:scenario-text}

\textbf{Scenario \#1: General-purpose technology}. \texttt{The private firm AI-X is developing a generative AI model capable of powering a hyper-realistic conversational video-chat agent. The system's user interface is very similar to common video conferencing softwares: the user logs in, describes the characteristics of the agent they would like to converse with, then a video chatbot exhibiting those characteristics appears on the screen and engages in an open-ended conversation with the user. 
You work at AI-X, and you have been tasked with completing an AIA for your team’s project (i.e., the video chatbot).}

\textbf{Scenario \#2: Hiring}. \texttt{The private firm AI-X has developed a generative AI model capable of powering a hyper-realistic conversational video-chat agent. The system's user interface is very similar to common video conferencing softwares: the system owner describes the characteristics of the agent they would like to create, the user logs in. then the video chatbot appears on the screen and engages in a conversation with the user according to the goals and characteristics specified by the system owner. AI-X is working on a product that fine-tune this agent to conduct initial screening interviews with job candidates.}

\subsection{Organizational roles}
For both scenarios, we presented students with identical text for the first two roles:

\texttt{1. ML scientist representing the team in charge of design and development}
\texttt{2. Product manager in charge of making the project economically successful}

The text for the third and final role varied depending on which scenario the group was assigned to:

(Scenario \#1: General purpose technology) \texttt{3. A member of the evaluation team who represents the interests of potential users.}

(Scenario \#2: Hiring) \texttt{3. A member of the evaluation team who represents the interests of potential job applicants. }

Students were assigned to organizational roles alphabetically by last name.

\subsection{Pre-Questionnaire}\label{apdx:pre-q}

Before beginning the activity, students individually completed a pre-questionnaire Google Form that was designed to take 5 minutes.

\begin{enumerate}
\item (multiple choice) Please select your team.
\item  (likert) On a scale of 1 to 5, how excited are you about the potential uses of foundational generative AI models (e.g., the product assigned to your team) in socially consequential domains? (1=not excited at all--5=very excited).
\item (likert) On a scale of 1 to 5, how concerned are you about the potential uses of foundational generative AI models (e.g., the product
assigned to your team) in socially consequential domains? (1=not concerned at all--5=very concerned).
\item (free response) Please outline the top three issues you believe the development team of the AI product assigned to your team should
address carefully before product release.
\item (multiple choice) Which one of the following broad categories of values do you believe is most urgent to address for your product? (A high-
level description of each value--taken from prior work--is provided below. Options are ordered randomly).
\begin{itemize}
    \item Fairness
    \item Safety
    \item Transparency
    \item Privacy
    \item Accountability \& governance
    \item Human autonomy \& agency
    \item Performance \& efficiency
\end{itemize}
\item (likert) On a scale of 1 to 5, what is the relative level of responsibility you imagine for ML experts (compared to other stakeholders)
to address the above matters? (1=very low at all--5=very high).
\item (free response) If you have any feedback for the teaching instructor or the research team about the questionnaire, please leave your
comments here.
\end{enumerate}

\subsection{Post-Questionnaire}\label{apdx:post-q}

After completing the activity, students individually completed a post-questionnaire Google Form that was designed to take no more than 10 minutes.  Note that all but two questions (Q2-3) which are new, are repeated verbatim from the pre-questionnaire.

\begin{enumerate}
\item (multiple choice) Please select your team.
\item (multiple choice) What was your individual role in your team?
\item (free response) What do you think are the major limitations of the AIA you completed for your team's product? 
\item  (likert) On a scale of 1 to 5, how excited are you about the potential uses of foundational generative AI models (e.g., the product assigned to your team) in socially consequential domains? (1=not excited at all--5=very excited).
\item (likert) On a scale of 1 to 5, how concerned are you about the potential uses of foundational generative AI models (e.g., the product
assigned to your team) in socially consequential domains? (1=not concerned at all--5=very concerned).
\item (free response) Please outline the top three issues you believe the development team of the AI product assigned to your team should
address carefully before product release.
\item (multiple choice) Which one of the following broad categories of values do you believe is most urgent to address for your product? (A high-
level description of each value--taken from prior work--is provided below. Options are ordered randomly).
\begin{itemize}
    \item Fairness
    \item Safety
    \item Transparency
    \item Privacy
    \item Accountability \& governance
    \item Human autonomy \& agency
    \item Performance \& efficiency
\end{itemize}
\item (likert) On a scale of 1 to 5, what is the relative level of responsibility you imagine for ML experts (compared to other stakeholders)
to address the above matters? (1=very low at all--5=very high).
\item (free response) If you have any feedback for the teaching instructor or the research team about the questionnaire, please leave your
comments here.
\end{enumerate}

\subsection{Value Category Definitions}\label{apdx:value-categories}

We provided students with a brief description of each of the seven value categories taken verbatim from prior work \citep{jakesch2022how}.

\begin{itemize}
    \item \textbf{Transparency}: A transparent AI system produces decisions that people can understand. Developers of transparent AI systems ensure, as far as possible, that users can get insight into why and how a system made a decision or inference.
    \item \textbf{Fairness}: A fair AI system treats all people equally. Developers of fair AI systems ensure, as far as possible, that the system does not reinforce biases or stereotypes. A fair system works equally well for everyone independent of their race, gender, sexual orientation, and ability.
    \item \textbf{Safety}: A safe AI system performs reliably and safely. Developers of safe AI systems implement strong safety measures. They anticipate and mitigate, as far as possible, physical, emotional, and psychological harms that the system might cause.
    \item \textbf{Accountability}: An accountable AI system has clear attributions of responsibilities and liability. Developers and operators of accountable AI systems are, as far as possible, held responsible for their impacts. An accountable system also implements mechanisms for appeal and recourse.
    \item \textbf{Privacy}: An AI system that respects people’s privacy implements strong privacy safeguards. Developers of privacy-preserving AI systems minimize, as far as possible, the collection of sensitive data and ensure that the AI system provides notice and asks for consent.
    \item \textbf{Human Autonomy \& Agency}: An AI system that respects people’s autonomy avoids reducing their agency. Developers of autonomy-preserving AI systems ensure, as far as possible, that the system provides choices to people and preserves or increases their control over their lives.
    \item \textbf{Performance \& Efficiency}: A high-performing AI system consistently produces good predictions, inferences or answers. Developers of high-performing AI systems ensure, as far as possible, that the system’s results are useful, accurate and produced with minimal delay.

\end{itemize}

\newpage
\section{Extended Results}\label{apdx:results-extended}

In this section, we present results from additional analyses excluded from the main text due to space constraints.

\newcommand{\mybreak}{%
  \par
  \nointerlineskip
  \cleaders\vbox to 5ex{%
    \vss
    \hbox to \textwidth{\hss\vrule width 0.25\textwidth height 0.2pt depth 0.2pt\hss}
    \vss
  }\vskip5ex
}

\subsection{Excitement \& Concern}

\begin{figure}[H]
\centering
\includegraphics[width=1\linewidth]{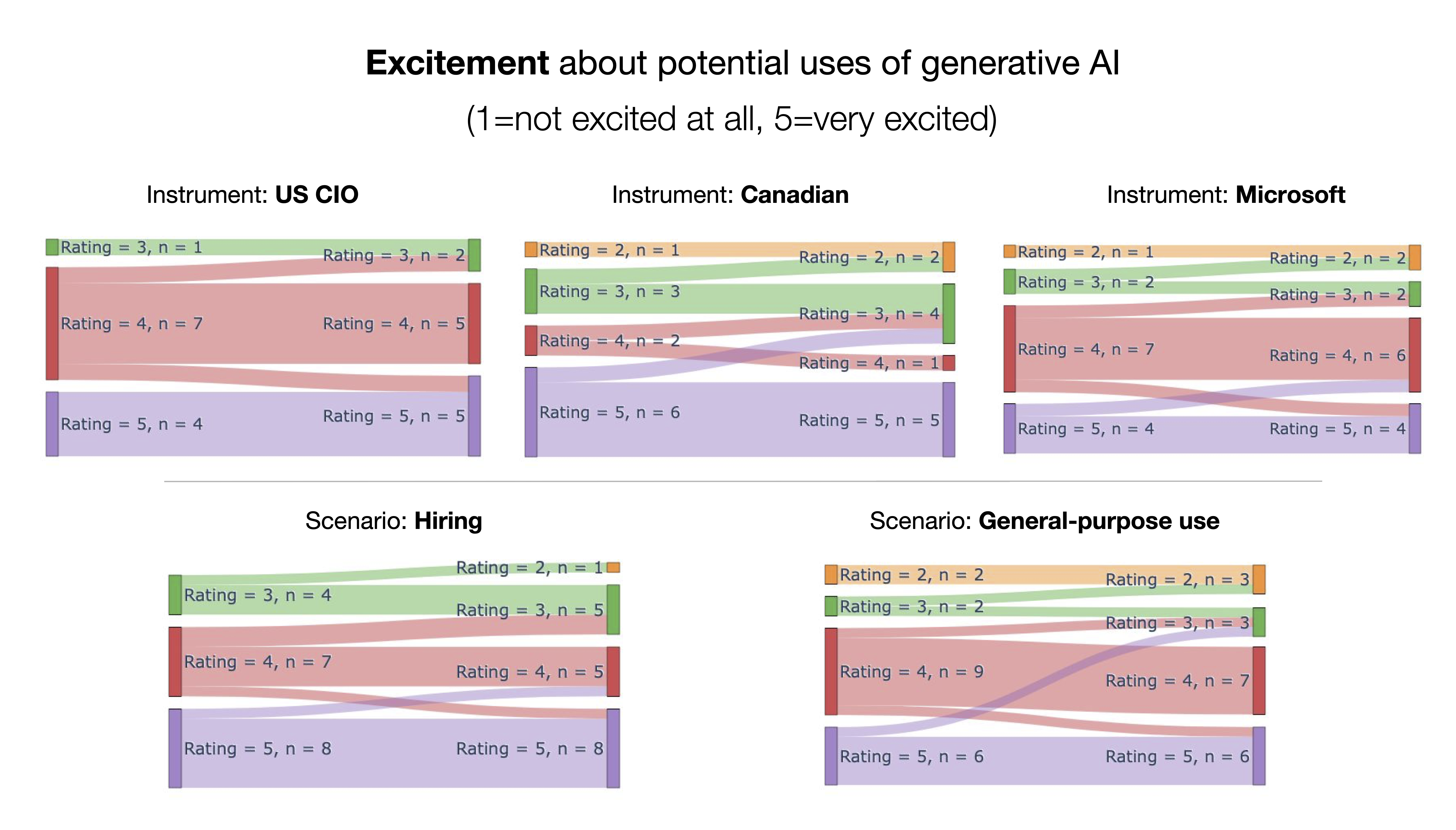}
\caption{Sankey plots visualizing changes in students' response to the question, ``\emph{On a scale of 1 to 5, how \textbf{excited} are you about the potential uses of foundational generative AI models (e.g., the product assigned to your team) in socially consequential domains? (1=not excited at all--5=very excited).}'' before (left) and after (right) completing the AIIA.  We group students' responses by the AIIA instrument (Top) and scenario (Bottom) that their team was assigned.}
\label{fig:sankey-excited}
\end{figure}

We report the average pre- and post-questionnaire excitement and concern ratings, stratified by (1) AIIA instrument and (2) scenario.  We visualize the distribution of individual students' responses before and after completing the AIA in Figures \ref{fig:sankey-excited} and \ref{fig:sankey-excited}.\\

\underline{AIIA Instrument - \textbf{Excitement} about potential uses of generative AI}:
\begin{itemize}
    \item \textbf{US CIO} ($n = 12$): $4.25 \rightarrow 4.25$ ($-0$)
    \begin{itemize}
        \item 2 out of 12 students changed their mind
        \item 1 student became less excited
    \end{itemize}
    \item \textbf{Canadian Treasury} ($n = 12$):  $4.08 \rightarrow 3.75$ ($-0.33$)
    \begin{itemize}
        \item 3 out of 12 students changed their mind
        \item 3 students became less excited
    \end{itemize}
    \item \textbf{Microsoft} ($n = 14$): $ 4.00 \rightarrow 3.86$ ($-0.14$)
    \begin{itemize}
        \item 4 out of 14 students changed their mind
        \item 3 students became less excited
    \end{itemize}
\end{itemize}

\underline{Scenario - \textbf{Excitement} about potential uses of generative AI}:
\begin{itemize}
    \item \textbf{Hiring} ($n = 19$): $4.21 \rightarrow 4.05$ ($-0.16$)
    \begin{itemize}
        \item 5 out of 19 students changed their mind
        \item 4 students became less excited
    \end{itemize}
    \item \textbf{General use} ($n = 19$): $4.0 \rightarrow 3.84$ ($-0.16$)
    \begin{itemize}
        \item 4 out of 19 students changed their mind
        \item 3 students became less excited
    \end{itemize}
\end{itemize}

\mybreak


\underline{AIIA Instrument - \textbf{Concern} about potential uses of generative AI}:
\begin{itemize}
    \item \textbf{US CIO} ($n = 12$): $4.08 \rightarrow 4.50$ ($+0.417$)

    \begin{itemize}
        \item 6 out of 12 students changed their mind
        \item 5 student became more concerned
    \end{itemize}
    \item \textbf{Canadian Treasury} ($n = 12$):  $4.08 \rightarrow 4.17$ ($+0.083$)
    \begin{itemize}
        \item 5 out of 12 students changed their mind
        \item 3 students became  more concerned
    \end{itemize}
    \item \textbf{Microsoft} ($n = 14$): $ 3.86 \rightarrow 4.29$ ($+0.43$)
    \begin{itemize}
        \item 8 out of 14 students changed their mind
        \item 7 students became  more concerned
    \end{itemize}
\end{itemize}

\underline{Scenario - \textbf{Concern} about potential uses of generative AI}:
\begin{itemize}
    \item \textbf{Hiring} ($n = 19$): $4.11 \rightarrow 4.26$ ($+0.16$)
    \begin{itemize}
        \item 8 out of 19 students changed their mind
        \item 5 students became more concerned
    \end{itemize}
    \item \textbf{General use} ($n = 19$): $3.89 \rightarrow 4.37$ ($+0.47$)
    \begin{itemize}
        \item 11 out of 19 students changed their mind
        \item 10 students became more concerned
    \end{itemize}
\end{itemize}

\subsection{Level of Responsibility}\label{apdx:resp-results}
We report the average pre- and post-questionnaire relative levels of responsibility stratified by condition, where a rating of $1$ is very low and $5$ is very high.

\underline{AIIA Instrument - Relative \textbf{responsibility} of ML experts}:
\begin{itemize}
    \item \textbf{US CIO} ($n = 12$): $4.42 \rightarrow 4.58$ ($+0.17$)

    \begin{itemize}
        \item 6 out of 12 students changed their mind
        \item 4 student reported increased responsibility
    \end{itemize}
    
    \item \textbf{Canadian Treasury} ($n = 12$):  $3.83 \rightarrow 4.25$ ($+0.42$)
    \begin{itemize}
        \item 9 out of 12 students changed their mind
        \item 7 students reported increased responsibility
    \end{itemize}
    \item \textbf{Microsoft} ($n = 14$): $4.4 \rightarrow 4.64$ ($+0.21$)
    \begin{itemize}
        \item 5 out of 14 students changed their mind
        \item 4 students reported increased responsibility
    \end{itemize}
\end{itemize}

\underline{Scenario - Relative \textbf{responsibility} of ML experts}:
\begin{itemize}
    \item \textbf{Hiring} ($n = 19$): $4.26 \rightarrow 4.63$ ($+0.37$)
    \begin{itemize}
        \item 10 out of 19 students changed their mind
        \item 8 students reported increased responsibility
    \end{itemize}
    \item \textbf{General use} ($n = 19$): $4.21 \rightarrow 4.37$ ($+0.16$)
    \begin{itemize}
        \item 10 out of 19 students changed their mind
        \item 7 students reported increased responsibility
    \end{itemize}
\end{itemize}

\newpage

\begin{figure}[H]
\centering
\includegraphics[width=1\linewidth]{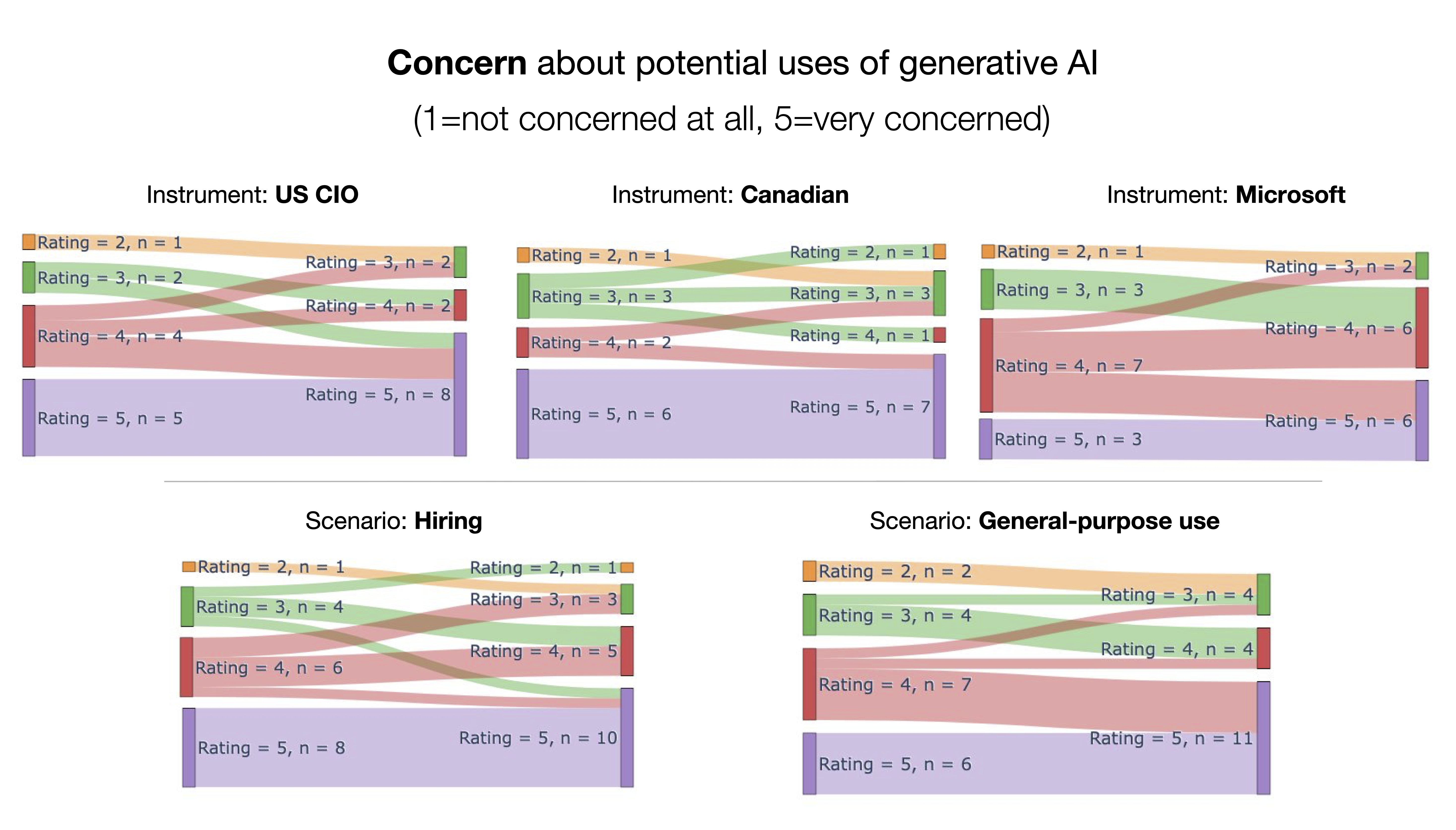}
\caption{Sankey plots visualizing changes in students' response to the question, ``\emph{On a scale of 1 to 5, how \textbf{concerned} are you about the potential uses of foundational generative AI models (e.g., the product
assigned to your team) in socially consequential domains? (1=not concerned at all--5=very concerned).}'' before (left) and after (right) completing the AIIA.  We group students' responses by the AIIA instrument (Top) and scenario (Bottom) that their team was assigned.}
\label{fig:sankey-concerned}
\end{figure}

\end{document}